\documentclass[showpacs,preprintnumbers,amsmath,amssymb,twocolumn,superscriptaddress,prb]{revtex4}
\usepackage{times}
\usepackage{graphicx}
\usepackage{amsfonts}
\usepackage{amsmath, amsthm, amssymb}
\usepackage{dsfont}
\usepackage{subfigure}

\newcommand{\vpf}{\mathbf{p}_\text{F}}

\newcommand{\e}[1]{\mathrm{e}^{#1}}

\newcommand{\gi}{g_\text{imp}}
\newcommand{\gs}{g_\text{sf}}

\newcommand{\eg}{\textit{e.g. }}
\newcommand{\etal}{\emph{et al. }}
\def\i{\mathrm{i}}

\begin{document}
\title[Proximity effect in ferromagnet/superconductor hybrids: from diffusive to ballistic motion]{Proximity effect in ferromagnet/superconductor hybrids: from diffusive to ballistic motion}
\author{Jacob Linder}
\affiliation{Department of Physics, Norwegian University of
Science and Technology, N-7491 Trondheim, Norway}
\author{Malek Zareyan}
\affiliation{Institute for Advanced Studies in Basic Sciences, 45195-1159, Zanjan, Iran}
\author{Asle Sudb{\o}}
\affiliation{Department of Physics, Norwegian University of
Science and Technology, N-7491 Trondheim, Norway}

\date{Received \today}
\begin{abstract}
We present an analytical study of the proximity effect in
ferromagnet/superconductor (F/S) heterostructures, allowing for an
arbitrary magnetic exchange energy as well as arbitrary impurity and spin-flip scattering rates within a quasiclassical approach. While previous
studies mainly have focused on the clean or dirty limits, our
results grant access to the regime of intermediate impurity
concentrations, thus allowing us to probe the crossover from the
clean to dirty limit. We find that in the crossover regime, all
possible symmetry correlations of the proximity-induced anomalous
Green's function are induced in the ferromagnet. We also point out that 
the local density of states oscillates spatially, not only for an F/S
bilayer, but also for a normal/superconductor (N/S) bilayer in the
diffusive limit, a fact which appears to have gone unnoticed in
the literature. Within the weak-proximity effect regime, we
present compact analytical expressions valid for arbitrary
exchange fields and impurity scattering rates for \textit{i)} the
local density of states in an F/S bilayer, \textit{ii)} the
Josephson current in an S/F/S junction, and \textit{iii)} the
critical temperature in an F/S/F multilayer. For all cases, we
study in particular the crossover regime between diffusive and
ballistic motion. Our results may be useful for analyzing
experimental data in cases when the dirty limit is not fully
reached, thus invalidating the use of the Usadel equation.
\end{abstract}

\maketitle

\section{Introduction}

The interest in ferromagnet/superconductor (F/S) heterostructures
has increased much during the last decade \cite{bergeretrmp,buzdinrmp,izyumov_review_02}.
This may probably be attributed to advances in experimental
fabrication/deposition techniques as well as intriguing theoretical 
predictions. The main hope is that future devices and applications 
will rely on manipulation of not only the electron charge but also 
its \textit{spin}. Based on this idea, a new research area known as 
superspintronics has emerged, aiming at utilization of charge and 
spin transport in ferromagnet/superconductor heterostructures. For 
instance, several authors have investigated the possibility of 
dissipationless currents of spin and charge in magnetically ordered
superconductors \cite{kulic, bergeretJOS, nogueira, shen,
linderPRL, asano, brydon, champelPRL}. A large
number of other studies related to spin degrees of freedom in
superconducting systems has also appeared in the literature
\cite{brataas,dani,taddei,linder08_2}.

\par
A considerable amount of attention has been devoted to the arguably most 
simple experimental laboratory where the interplay between ferromagnetism 
and superconductivity may be studied, namely a F/S bilayer. The two long-range 
order phenomena mix close to the interface, giving rise to interesting effects 
both from a basic physics perspective and in terms of potential applications. 
These effects include induction of unusual superconducting symmetry
correlations and a highly non-monotonic behaviour of various physical 
quantities on the size of the system. The latter is a result of the 
non-uniform superconducting correlations that are induced in the 
ferromagnetic layer by means of the proximity effect.

\par
As a natural extension of the F/S bilayer, there has also been
much focus on S/F/S systems and F/S/F systems, where the influence
of ferromagnetism on the Josephson current and the critical
temperature has been studied, respectively. The large majority of
works related to these systems assumed that the diffusive limit
was reached. In this case, elastic scattering on impurities
renders the Green's function to be isotropic in space, while it
may still retain a complicated spin-structure. From an
experimental point of view, the diffusive regime is certainly
relevant, but there are nevertheless some complications. One point
bears upon the theoretical framework used to study the physics in
the diffusive regime. The quasiclassical Usadel \cite{usadel}
equation is widely employed to study the proximity effect in F/S
heterostructures, and is valid under two main assumptions.
Firstly, that the Fermi energy is much larger than any other
energy scale and the essential physics is governed by fermions at
Fermi level, and secondly, that the inverse impurity scattering
rate is much larger than any other energy scale except for the
Fermi energy. For strong ferromagnets such as Co or Ni,
the second condition may be violated. In that case, one must
revert to the more general Eilenberger \cite{eilenberger} equation,
which is only subject to the first condition.

\par
The Eilenberger equation is more complicated to solve analytically
than the Usadel equation, although some special limits permit
fairly simple analytical expressions. Let $h$ denote the
exchange-energy of the ferromagnet while $\tau_\text{imp}$ denotes
the inverse impurity scattering rate. The Usadel equation is then
obtained from the Eilenberger equation by demanding
$h\tau_\text{imp} \ll 1$, while the case of a strong and clean
ferromagnet is obtained in the limit $h\tau_\text{imp} \gg 1$. We 
assume that $h\gg\Delta$ is fulfilled. In Ref. \cite{bergeret02}, some 
aspects of the DOS in F/S heterostructures were considered to leading 
order in the parameter $(h\tau_\text{imp})^{-1}$, corresponding to a 
strong ferromagnet which falls outside the range of applicability of the
Usadel equation. In Ref. \cite{bergeretstrong}, the Josephson
current in an S/F/S structure was also investigated for the case
of a strong ferromagnet, $h\tau_\text{imp} \gg 1$. Some authors
have also considered F/S heterostructures where the impurity
scattering rate was disregarded or assumed to be small,
corresponding to the ballistic regime. \cite{zareyan_cleanlimit,
eschrig_pure,valls,linder_pure, buzdin_jetp_82, fogelstrom_prb_00, belzig_jetp_01, barash_prb_02}
\par
Although the agreement between theory and experiment in this
research area has proven to be satisfactory in many cases, there
are still discrepancies to be accounted for. For instance, the
Usadel equation has failed to account quantitatively for the
critical temperature in F/S/F spin-valves. Furthermore, anomalous
features in the DOS for a very thin F/S bilayer that could
not be accounted for even qualitatively, were reported in Ref.
\cite{sangiorgio}. Moreover, the Usadel equation approach fails
from the start when addressing systems with strong ferromagnets.

\par
\begin{figure}[htb]
\begin{center}
\scalebox{0.30}{
\includegraphics[width=19.0cm,clip]{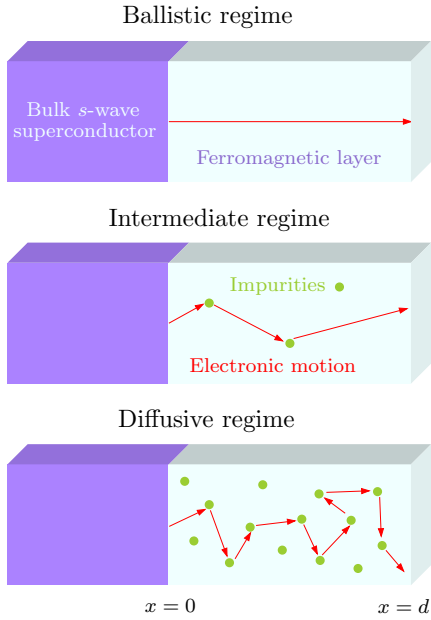}}
\end{center}
\caption{(color online) Overview of the superconductor/ferromagnet
heterostructure we will study in this paper. We take into account
an arbitrary strength of the exchange field as well as an arbitrary rate of non-magnetic and magnetic scattering within a quasiclassical approach. }
\label{fig:model}
\end{figure}

All of this points to the need of taking the role of impurity
scattering more seriously. In this paper, we aim at doing
precisely so by solving the Eilenberger equation analytically and
studying the \textit{crossover} regime between ballistic and
diffusive motion (see Fig. \ref{fig:model}). To illustrate how
various physical quantities behave in this crossover regime, we
study \textit{i)} the local density of states in an F/S bilayer,
\textit{ii)} the Josephson current in an S/F/S junction, and
\textit{iii)} the critical temperature in an F/S/F multilayer for
\textit{arbitrary} values of $h$ and $\tau$ (within the
quasiclassical approach). In each case, we present compact
analytical formula to facilitate comparison to experimental data
in cases where the diffusive limit may not be fully warranted or
where strong ferromagnets are involved.
\par
This paper is organized as follows. In Sec. \ref{sec:theory}, we
establish the theoretical framework which is employed in this
work. In Sec. \ref{sec:results}, we present our main results with
belonging discussion: the DOS of an F/S bilayer in Sec.
\ref{sec:dos}, the Josephson current in an S/F/S multilayer in
Sec. \ref{sec:jos}, and finally the critical temperature in an
F/S/F multilayer in Sec. \ref{sec:tc}. Eqs. (\ref{eq:DOS}), (\ref{eq:josmain}), 
and (\ref{eq:Tc}) are the main analytical results of this work. We conclude in Sec.
\ref{sec:summary}. Below, we will use boldface notation for vectors, $\underline{\ldots}$ for $2\times2$ matrices, and $\hat{\ldots}$ for $4\times4$ matrices. The reader may consult the Appendix for a definition of the generalized Pauli-matrices we employ in this paper.

\par
\section{Theoretical formulation}\label{sec:theory}
The Eilenberger equation reads \cite{eilenberger}
\begin{align}\label{eq:eilenberger}
\i \mathbf{v}_F \cdot \nabla \hat{g} + [\varepsilon\hat{\rho}_3 + \hat{M} - \hat{V}_\text{imp} - \hat{S}_\text{flip} + \hat{\Delta}, \hat{g}] = 0,
\end{align}
where $\hat{g} \equiv \hat{g}^\text{R}(\mathbf{R},\varepsilon,
\mathbf{p}_F)$ is the retarded part of the Green's function. Here, $\varepsilon$ is the quasiparticle energy, $\mathbf{R}$ is the center-of-mass coordinate, and $\mathbf{p}_F$ ($\mathbf{v}_F$) is the Fermi momentum (velocity) vector. The
self-energies that enter Eq. (\ref{eq:eilenberger}) are the
magnetic exchange energy $\hat{M} =
h\text{diag}\{\underline{\tau_3},\underline{\tau_3}\}$, the
impurity scattering $\hat{V}_\text{imp} =
-[\i/(2\tau_\text{imp})]\langle \hat{g}\rangle$, the (uniaxial)
spin-flip scattering $\hat{S}_\text{flip} =
-[\i/(2\tau_\text{flip})] \hat{\rho}_3\langle \hat{g} \rangle
\hat{\rho}_3$, and the superconducting order parameter
\begin{align}
\hat{\Delta} = \begin{pmatrix}
\underline{0} & \i\underline{\tau_2}\Delta \notag\\
\i\underline{\tau_2}\Delta^* & \underline{0} \notag\\
\end{pmatrix}.
\end{align}
All matrices used above ($\hat{\rho}_i, \underline{\tau_i}$) are defined in the Appendix [Eq. (\ref{eq:matricesapp})]. 
The brackets $\langle\ldots\rangle$ denote an angular average over
the Fermi surface. Also, $h$ is the exchange splitting while
$\tau_\text{imp(flip)}$ is the scattering time associated with
impurity (spin-flip) scattering. We may conveniently rewrite Eq.
(\ref{eq:eilenberger}) as:
\begin{align}
\i \mathbf{v}_F \cdot &\nabla \underline{g_\sigma} + [(\varepsilon+\sigma h)\underline{\tau_3} + \sigma\underline{\Delta} + \frac{\i}{2\tau_\text{imp}}\langle \underline{g_\sigma}\rangle \notag\\
&+ \frac{\i}{2\tau_\text{sf}} \underline{\tau_3} \langle \underline{g_\sigma}\rangle \underline{\tau_3}, \underline{g_\sigma}] = 0,\; \sigma=\uparrow,\downarrow=\pm1
\end{align}
where the superconducting order parameter matrix $\underline{\Delta}$ reads
\begin{equation}
\underline{\Delta} = \begin{pmatrix}
0 & \Delta\\
-\Delta^* & 0\\
\end{pmatrix},\; \Delta=\Delta_0\e{\i\chi},
\end{equation}
upon letting $\chi$ denote the phase corresponding to the globally broken U(1) symmetry in the superconducting state. 
The brackets $\langle \ldots \rangle$ denote angular averaging over the Fermi surface.
We employ the Ricatti parametrization \cite{schopohl} of the Green's function:
\begin{equation}
\underline{g_\sigma} = \mathcal{N}_\sigma\begin{pmatrix}
1 - a_\sigma b_\sigma & 2a_\sigma \\
2b_\sigma & -1 + a_\sigma b_\sigma\\
\end{pmatrix},\; \mathcal{N}_\sigma = (1+a_\sigma b_\sigma)^{-1}.
\end{equation}
Here, $a_\sigma$ and $b_\sigma$ are two unknown functions used to parametrize the 
Green's functions. They will be determined by solving the Eilenberger equation with 
appropriate boundary conditions. A general treatment of the Eilenberger equation 
calls for a numerical solution. In the case of a weak proximity effect, however, 
the Eilenberger equation may be linearized in the anomalous part of the Green's 
function which permits an analytical approach. The assumption of a weak proximity effect
corresponds mathematically to a scenario where higher order terms of $\{a_\sigma,b_\sigma\}$ are 
disregarded in the Eilenberger equation, i.e. one assumes that $|a_\sigma|\ll1$, $|b_\sigma|\ll1$. 
In an experimental situation, a weak proximity effect in F/S heterostructures may be expected whenever
the tunneling limit is reached and the number of conducting channels at the interface is low. Also, 
assuming a superconducting reservoir, the proximity effect becomes weaker in magnitude upon
increasing the thickness of the ferromagnetic layer.
\par
The spatial depletion of the superconducting order parameter near
the S/F interface will be disregarded. This is an excellent
approximation in the corresponding low-transparency regime, which
will be considered throughout this paper except for in Sec. \ref{sec:tc}, where this issue is discussed further. At the S/F interface ($x=0$) we use
Zaitsev's boundary conditions. Define the symmetric and
antisymmetric part of the Green's function as
\begin{align}
\underline{\mathcal{S}_{\sigma,i}} = \frac{1}{2}(\underline{g_{\sigma,i}}^+ + \underline{g_{\sigma,i}}^-), \; \underline{\mathcal{A}_{\sigma,i}} = \frac{1}{2}(\underline{g_{\sigma,i}}^+ - \underline{g_{\sigma,i}}^-),
\end{align}
where the $\pm$ superscript on the Green's function denotes
right/left-going quasiparticle excitations and the subscript $i$
denotes the ferromagnetic or superconducting region. The first of
Zaitsev's boundary conditions \cite{zaitsev} demands continuity of
the antisymmetric part $\underline{\mathcal{A}_{\sigma,i}}$ of the
Green's function. The second one relates the Green's functions in
the ferromagnetic and superconducting regions to the interface
transparency. We obtain
\begin{align}\label{eq:zaitsev}
\underline{\mathcal{A}_{\sigma,F}}[\mathcal{R}(1-\underline{\mathcal{A}_{\sigma,F}}^2) &+ \frac{\mathcal{T}}{4} (\underline{\mathcal{S}_{\sigma,S}}-\underline{\mathcal{S}_{\sigma,F}})^2] \notag\\
& = \frac{\mathcal{T}}{4} [\underline{\mathcal{S}_{\sigma,F}}, \underline{\mathcal{S}_{\sigma,S}}]_-,
\end{align}
where $\mathcal{R}$ and $\mathcal{T}$ are the reflection and transmission
coefficients satisfying $\mathcal{R}+\mathcal{T}=1$, and $[\ldots]_-$ denotes a
commutator. High and low transparency interfaces correspond to
$\mathcal{T}\simeq 1$ and $\mathcal{T}\ll 1$, respectively. Although Eq.
(\ref{eq:zaitsev}) is expressed rather compactly, a general
solution for arbitrary $\mathcal{T}$ and $\mathcal{R}$ is very hard to obtain. In the
experimentally relevant situation, one may assume that $\mathcal{T} \ll \mathcal{R}$.
For a low-transparency barrier and a weak proximity effect, Eq.
(\ref{eq:zaitsev}) simplifies greatly to
\begin{align}\label{eq:zaitsev_simple}
\underline{\mathcal{A}_{\sigma,F}}|_{x=0} = \gamma [\mathcal{S}_{\sigma,F}, \mathcal{S}_{\sigma,S}]_-|_{x=0},
\end{align}
where $\gamma = \mathcal{T}/(4\mathcal{R})$ is a measure of the barrier transparency.
At the end of the ferromagnetic layer, we demand
$\underline{\mathcal{A}_{\sigma,F}}|_{x=d} = \underline{0}$.
\par
We consider here an effective one-dimensional calculation, which should provide 
sound results due to the isotropic nature of the ferromagnetic and
superconducting order parameters. We do not expect any qualitative differences from a two-dimensional
or three-dimensional model, since the superconducting gap and the magnetic exchange field do not depend
on the quasiparticle momenta, and since there are no surface-bound states \cite{hu_prl_94} at the interfaces of the systems we consider. Thus, it should be possible to capture the essential physics by studying an effective one-dimensional model, which permits us to proceed analytically. This point of view is supported by the fact that, as seen later in this work, we reproduce in limiting cases previous results obtained in the literature which employed a two-dimensional calculation.
\par
Under the assumption of a weak
proximity effect, the Eilenberger equations in the ferromagnetic
region take the form:
\begin{align}\label{eq:case1}
\alpha\i v_F \partial_x a_\sigma + 2a_\sigma(\varepsilon+\sigma h) &+ \frac{\i}{2\tau_\text{imp}} (a_\sigma^\alpha - a_\sigma^{-\alpha}) \notag\\
&+ \frac{\i}{2\tau_\text{sf}} (3a_\sigma^\alpha + a_\sigma^{-\alpha}) = 0\notag\\
\alpha\i v_F \partial_x b_\sigma - 2b_\sigma(\varepsilon+\sigma h) &- \frac{\i}{2\tau_\text{imp}} (b_\sigma^\alpha - b_\sigma^{-\alpha}) \notag\\
&- \frac{\i}{2\tau_\text{sf}} (3b_\sigma^\alpha + b_\sigma^{-\alpha}) = 0,
\end{align}
where $\alpha=\pm$ denotes
right- and left-going quasiparticles, respectively. It is necessary to take into account the direction of the quasiparticles at Fermi level due to the term $\mathbf{v}_F \cdot \nabla \hat{g}$ in Eq. (\ref{eq:eilenberger}). Thus, $\sigma$ denotes the spin direction while $\alpha$ denotes the direction of motion in $a_\sigma^\alpha$ and likewise for $b_\sigma^\alpha$. The impurity and spin-flip scattering self-energies enter Eq. (\ref{eq:case1}) by means of the matrices $\hat{V}_\text{imp}$ and $\hat{S}_\text{flip}$ in Eq. (\ref{eq:eilenberger}), which both depend on the Fermi-surface averaged Green's function. For a weak proximity effect, we have
\begin{align}
\langle \underline{g_\sigma}\rangle &= \begin{pmatrix}
1 & a_\sigma^+ + a_\sigma^- \\
b_\sigma^+ + b_\sigma^- & -1\\
\end{pmatrix}.
\end{align} 
For a bulk ferromagnet, the solution is $a_\sigma^\pm=b_\sigma^\pm=0$.
\par
In Ref. \cite{bergeret02}, the DOS in a S/F bilayer was studied by
neglectinb both spin-flip scattering ($\tau_\text{sf}\to\infty$)
and the coupling term between the right- and left-going
excitations in Eq. (\ref{eq:case1}). In this case, one finds that
Eq. (\ref{eq:case1}) reduces to
\begin{align}
\pm \i v_F\partial_x a_\sigma^\pm &+ [2(\varepsilon+\sigma h) + \frac{\i}{2\tau_\text{imp}}]a_\sigma^\pm = 0,\notag\\
\pm \i v_F\partial_x b_\sigma^\pm &- [2(\varepsilon+\sigma h) + \frac{\i}{2\tau_\text{imp}}]b_\sigma^\pm = 0.
\end{align}
The decaying solution for $x\to\infty$ of the above equations reads
\begin{align}
a_\sigma^+ = k_{a\sigma} \exp[-\kappa_\sigma x/l],\; b_\sigma^- = k_{b\sigma} \exp[-\kappa x/l],\notag\\
\kappa_\sigma = 1 - 2\i(\varepsilon+\sigma h)\tau_\text{imp},\; l = v_F\tau_\text{imp}.
\end{align}
while $a_\sigma^-=b_\sigma^+=0$. Above, $k_{a\sigma}$ and
$k_{b\sigma}$ are constants to be determined from the boundary
condition at $x=0$, and the structure of the Green's function
becomes
\begin{align}
\mathcal{S}_{\sigma,F} =
\begin{pmatrix}
1 & a_\sigma^+ \\
b_\sigma^- & -1 \\
\end{pmatrix},\;
\mathcal{A}_{\sigma,F} =
\begin{pmatrix}
0 & a_\sigma^+ \\
-b_\sigma^- & 0 \\
\end{pmatrix}.\;
\end{align}
This shows how the decay length of the proximity-induced anomalous
Green's function in the ferromagnet is governed by the mean free
path $l$, and that it is independent of the exchange field in this
main approximation. We now present a more rigorous solution by
fully taking into account the coupling-term in Eqs.
(\ref{eq:case1}). To solve this problem, we note that Eqs.
(\ref{eq:case1}) may be written as a matrix differential equation:
\begin{align}
\partial_x\mathbf{a}_\sigma &= \underline{M_{a\sigma}} \mathbf{a}_\sigma,\; \mathbf{a}_\sigma = [a_\sigma^+ a_\sigma^-]^\mathrm{T},\notag\\
&\underline{M_{a\sigma}} = \frac{1}{v_F}
\begin{pmatrix}
r_\sigma & g \\
-g & -r_\sigma \\
\end{pmatrix},
\end{align}
where $\mathrm{T}$ denotes matrix transpose and we have defined the auxiliary quantities
\begin{align}
r_\sigma = 2\i(\varepsilon+\sigma h) - (\gi + 3\gs)/2,\notag\\
g = (\gi - \gs)/2,\; g_\text{imp(sf)} \equiv \tau_\text{imp(sf)}^{-1}.
\end{align}
Diagonalizing $\underline{M_\sigma}$ according to $\underline{D_\sigma} = \underline{P_\sigma}^{-1}\underline{M_\sigma}\underline{P_\sigma}$, we obtain the trivial set of decoupled differential equations
\begin{align}
\partial_x \mathbf{\tilde{a}}_\sigma = \underline{D_\sigma} \mathbf{\tilde{a}}_\sigma,\; \mathbf{\tilde{a}}_\sigma = \underline{P_\sigma}^{-1}\mathbf{a}_\sigma.
\end{align}
From the above, we find that
\begin{align}
\tilde{a}_\sigma^\pm = C_{a,\sigma}^\pm \e{\pm\lambda_\sigma x},\; \lambda_\sigma = v_F^{-1}\sqrt{r_\sigma^2- g^2},
\end{align}
while the diagonalization matrix $\underline{P_\sigma}$ reads
\begin{align}\label{eq:P}
\underline{P_\sigma} &= \begin{pmatrix}
p_{1\sigma} & p_{2\sigma} \\
p_{2\sigma} & p_{1\sigma}\\
\end{pmatrix},\; G_\sigma = g/(v_F\lambda_\sigma + r_\sigma),\notag\\
p_{1\sigma} &= N_\sigma,\; p_{2\sigma} = -N_\sigma G_\sigma,\; N_\sigma = (1 + |G_\sigma|^2)^{-1/2}.\notag\\
\end{align}
\par
In the superconducting region, we employ the bulk solution under the assumption that the interface transparency is low and that the ferromagnetic layer is much more disordered than the superconductor \cite{bergeretrmp}. In this main approximation, we may employ the bulk
solution of the Green's function in the superconductor:
\begin{align}
\underline{g_\sigma^\pm} &= \begin{pmatrix}
c(\theta) & \sigma s(\theta) \notag\\
-\sigma s(\theta) & -c(\theta) \notag\\
\end{pmatrix},
\end{align}
with the definitions $c(\theta) =\cosh(\theta)$, 
$s(\theta) = \sinh(\theta)$, $\theta = \text{atanh}(\Delta/\varepsilon)$. Once the expression for the Green's function in the ferromagnet
has been obtained, one may calculate various physical quantities
of interest. By approximating $\underline{\mathcal{S}_{\sigma,F}}
\simeq \underline{\tau_3}$ in Eq. (\ref{eq:zaitsev_simple}) in
accordance with a weak proximity effect, we obtain for the case
where the impurity-scattering coupling between the
Ricatti-equations is ignored:
\begin{align}
\underline{g_{\sigma,F}^\pm} = \underline{\tau_3} + &2\gamma\sigma s(\theta)\exp(-\kappa_\sigma x/l)(\underline{\tau_1} \pm \i\underline{\tau_2}),
\end{align}
which is precisely the result of Ref. \cite{bergeret02} for two
semi-infinite superconducting and ferromagnetic layers in contact.
When the coupling is properly taken into account, in addition to
the vacuum boundary condition at $x=d$, we find that
\begin{align}\label{eq:geilen}
\underline{g_{\sigma,F}^\pm} &= \begin{pmatrix}
1 & 2a_\sigma^\pm \\
2b_\sigma^\pm & - 1 \\
\end{pmatrix},\; \text{ upon defining }\notag\\
\text{ } \notag\\
a_\sigma^\pm &= p_\sigma^\pm C_{1\sigma}(\lambda_\sigma) + p_\sigma^\mp C_{2\sigma}(\lambda_\sigma), \notag\\
b_\sigma^\pm &= p_\sigma^\pm C_{1\sigma}(-\lambda_\sigma) + p_\sigma^\mp C_{2\sigma}(-\lambda_\sigma), \notag\\
\text{ } \notag\\
C_{1\sigma} &= \frac{2\gamma\sigma s(\theta)\e{\lambda_\sigma x}}{p_\sigma^+ - p_\sigma^-}\Big[1 - \frac{\e{\lambda_\sigma d}}{2\sinh(\lambda_\sigma d)} \Big], \notag\\
C_{2\sigma} &= -\frac{\gamma\sigma s(\theta) \e{\lambda_\sigma (d-x)}}{(p_\sigma^+-p_\sigma^-)\sinh(\lambda_\sigma d)},
\end{align}
and $p_\sigma^\pm = p_{1,2\sigma}$. Note that in the diffusive
limit where $\gi \gg \{h,\varepsilon,\Delta_0, \gs\}$, one would
expect that the distinction between right-going and left-going
particles is removed, such that $\underline{g_{\sigma,F}}^+ =
\underline{g_{\sigma,F}}^-$. This is easily shown by exploiting
\begin{align}
\lim_{\gi \gg \{h,\varepsilon,\Delta_0, \gs\}} (v_F\lambda_\sigma+r_\sigma) = -\gi/2,
\end{align}
as seen from the previous equations. We also want to compare the
results for $\gi \gg \{h,\varepsilon,\Delta_0,\gs\}$ with those
obtained when using the linearized Usadel equation. The Usadel
equation in a diffusive ferromagnet then reads
\begin{align}
D\partial_x^2 f_\pm + 2\i(\varepsilon  + \i\gs \pm h)f_\pm = 0,
\end{align}
where $f_\pm = f_t \pm f_s$ and $f_t$ is the odd-frequency triplet
anomalous Green's function while $f_s$ is the even-frequency
singlet anomalous Green's function (both are isotropic in momentum
space). We obtain that the only physically acceptable (decaying
for $x\to\infty$) solution is
\begin{align}\label{eq:f1}
f_+ &= f_0 \e{\i k_+x} \text{ if } \varepsilon > 0,\; f_- = f_0 \e{-\i k_- x} \text{ if } \varepsilon < 0,\notag\\
&k_\pm = \sqrt{2\i(\varepsilon + \i\gs \pm h)/D},
\end{align}
where $f_0$ is a constant to be determined from the boundary
conditions. Above, $D$ is the diffusion constant. For consistency,
we should be able to obtain the same decaying solution from Eq.
(\ref{eq:geilen}) when $\gi \gg \{h,\varepsilon,\Delta_0,\gs\}$.
Focusing on the wavevector, we see that in this limit:
\begin{align}\label{eq:lambda}
\lambda_\sigma &\to v_F^{-1}\sqrt{-2\i(\varepsilon+\sigma h)\gi +2\gi\gs}\notag\\
&= \sqrt{-2\i(\varepsilon +\sigma h+ \i\gs)/D},
\end{align}
where $D = v_F^2\tau_\text{imp}$ is the diffusion constant in one
dimension (in three dimensions, $D=v_F^2\tau_\text{imp}/3$). Eq.
(\ref{eq:lambda}) is then consistent with the form of Eq.
(\ref{eq:f1}).

\par
With a complete description of the behaviour of the Green's
function in the ferromagnetic region, we now investigate the
influence of the proximity effect on the local density of states
(LDOS), and also study the singlet and triplet superconducting
order parameters induced in the ferromagnet. The normalized LDOS
as obtained from the solution of the Eilenberger equation may be
written as
\begin{align}
N(\varepsilon,x) &= \frac{1}{2} \sum_\sigma \langle \text{Re}\{ 1 + 4a_\sigma(\varepsilon,x)b_\sigma(\varepsilon,x) \} \rangle
\end{align}
for a weak proximity effect. In the normal state, the normalized
DOS is $N_0=1$. Inserting the expressions for $a_\sigma^\pm$ and
$b_\sigma^\pm$ into the above equation yields
\begin{widetext}
\begin{align}\label{eq:DOS}
N&(x,\varepsilon) = 1 - \text{Re}\Bigg\{\sum_\sigma \frac{2\gamma^2s^2(\theta)}{\sinh(\lambda_\sigma d) (1+G_\sigma)^2}\times \Bigg[1 + G_\sigma^2 -2G_\sigma\Bigg(2\sinh(\lambda_\sigma d - 2\lambda_\sigma x) + \frac{\cosh(2\lambda_\sigma x)}{\sinh(\lambda_\sigma d)}\Bigg)\Bigg]\Bigg\}
\end{align}
\end{widetext}
Eq. (\ref{eq:DOS}) is the first of our three main analytical
results in this work. Within the weak proximity effect regime, it
provides a general expression for the DOS, taking into account an
arbitrary exchange field and impurity scattering rate. As seen,
the correction to the normal state DOS $N_0=1$ is zero for a
vanishing interface transparency $(\gamma=0)$. While the weak
proximity restriction only allows access to variations from the
normal-state of DOS of around 10$\%$, this seems to be sufficient
for the experimentally relevant situation. For instance, the
deviation from the normal-state DOS due to the superconducting
proximity effect was of order 1$\%$ in Ref.~\onlinecite{kontos}.

\par
In order to study the superconducting correlations inside the
ferromagnetic region, first note that the full structure of the
retarded Green's function is
\begin{align}
\hat{g}^\text{R} &= \begin{pmatrix}
\underline{g} & \underline{f} \\
-\underline{\tilde{f}} & -\underline{\tilde{g}} \\
\end{pmatrix},
\end{align}
where the spin-structure reads
\begin{align}
\underline{f} = \begin{pmatrix}
f_{\uparrow\uparrow} & f_{\uparrow\downarrow}\\
f_{\downarrow\uparrow} & f_{\downarrow\downarrow}\\
\end{pmatrix},
\end{align}
and we have defined $f_{\alpha\beta} = f_{\alpha\beta}(\vpf,\varepsilon,x)$ and $\tilde{f}(\vpf,\varepsilon,x) = f(-\vpf,-\varepsilon,x)^*$. From the Ricatti-parametrization, we may define the different symmetry-components of the anomalous Green's functions as follows:
\begin{align}\label{eq:comp}
f_\text{ESE} &= \sum_\sigma \sigma (a_\sigma^+ + a_\sigma^-),\; f_\text{OSO} = \sum_\sigma \sigma (a_\sigma^+ - a_\sigma^-),\notag\\
f_\text{ETO} &= \sum_\sigma (a_\sigma^+ - a_\sigma^-),\; f_\text{OTE} = \sum_\sigma (a_\sigma^+ + a_\sigma^-).
\end{align}
Here, the abbreviations are explained in Tab. \ref{tab:symmetry}.
Note that in the general case of finite $h$ and $\tau_\text{imp}$,
\textit{all possible symmetry components} of the anomalous Green's
function are induced in the non-superconducting region. In the
case of $h=0$, one may confirm from Eq. (\ref{eq:geilen}) that
$a_\sigma^\pm \to \sigma a^\pm$, where $a^\pm$ is independent of
$\sigma$, such that $f_\text{OTE} = f_\text{ETO} = 0$. Physically,
the induction of other symmetry components than $f_\text{ESE}$,
corresponding to the bulk superconductor, may be explained as
follows \cite{tanakaPRL,yokoyamaPRL}. In a normal
metal/superconductor junction, the translational symmetry is broken at the
interface separating the two regions. This causes even-parity and
odd-parity components of the Green's function to mix near the
interface. Since the Pauli-principle must be satisfied at all
times, a change in the parity-symmetry of the Green's function
must be accompanied by a change in either spin- or
frequency-symmetry. In the absence of an exchange field, nothing
breaks the spin symmetry, such that only the frequency-symmetry
may be altered indirectly by the broken translational symmetry.
However, if the spin-symmetry is also broken by replacing the
normal metal with a ferromagnet, the spin-symmetry of the Green's
function may also be altered. These considerations are summarized
in Tab. \ref{tab:symmetry}. The possibility of a \textit{bulk}
odd-frequency superconducting state was discussed in Refs.
\cite{berezinskii,abrahams}, and there has very recently been made
some predictions concerning characteristic transport properties of
such a bulk odd-frequency superconducting state.
\cite{tanakaPRL,oddJOS,fominovodd,oddprb}.
\begin{widetext}
\text{ }
\begin{table}[h!]
\centering{ \caption{Proximity-induced anomalous Green's functions
in a normal metal in contact with a conventional
BCS-superconductor, which has an even-frequency spin-singlet
even-parity symmetry. Below, the quasiballistic limit regime is
characterized by a vanishing or small value of the impurity
scattering rate, while the diffusive limit is characterized by an
impurity scattering rate which dominates all other energy scales
in the problem (except for the Fermi energy). }
    \label{tab:symmetry}
    \vspace{0.15in}
    \begin{tabular}{ccccc}
         \hline
         \hline
         Symmetry   \hspace{0.1in}  & $h\neq0$, quasiballistic \hspace{0.1in} & $h=0$, quasiballistic \hspace{0.1in} & $h\neq0$, diffusive\hspace{0.1in} & $h=0$, diffusive  \\
         \hline
         Even-frequency spin-singlet even-parity (ESE)  \hspace{0.1in} & $\surd$    \hspace{0.1in} & $\surd$    \hspace{0.1in} & $\surd$ \hspace{0.1in} & $\surd$ \\
         Odd-frequency spin-singlet odd-parity (OSO)    \hspace{0.1in} & $\surd$    \hspace{0.1in} & $\surd$    \hspace{0.1in} & - \hspace{0.1in} & -\\
         Even-frequency spin-triplet odd-parity (ETO)   \hspace{0.1in}   & $\surd$  \hspace{0.1in} & -  \hspace{0.1in} & - \hspace{0.1in} & -\\
         Odd-frequency spin-triplet even-parity (OTE)  \hspace{0.1in} & $\surd$ \hspace{0.1in} & -  \hspace{0.1in} & $\surd$ \hspace{0.1in} & -\\
         \hline
         \hline
    \end{tabular}}
\end{table}
\end{widetext}

\section{Results and discussion}\label{sec:results}

\subsection{Anomalous Green's functions}\label{sec:anomalous}
\begin{figure}[htb]
\begin{center}
\scalebox{0.45}{
\includegraphics[width=19.0cm,clip]{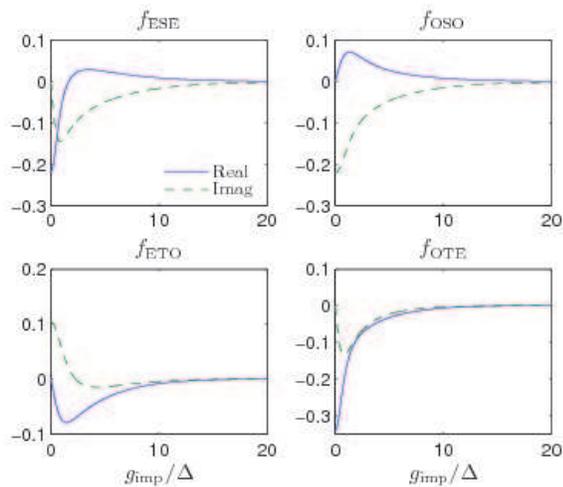}}
\end{center}
\caption{(color online) Plot of the proximity-induced anomalous
Green's functions in the middle of the ferromagnetic region
$(x/d=0.5)$ using $h/\Delta=15$ and $\varepsilon/\Delta_0=0.5$.}
\label{fig:hneq0}
\end{figure}
The linearized Eilenberger equations allow us to study the direct
crossover from the diffusive to the ballistic regime of
quasiparticle transport, and hence dependence of the different
symmetry components on the impurity scattering. In the
experimental situation, one usually probes the DOS at the F/I
interface $x=d$, although it in principle is possible to obtain a
spatially resolved DOS in the entire ferromagnetic region by using
local scanning tunneling microscopy (STM)-measurements. Let us
first focus on $x=d$ and consider the ballistic limit in which
case simple and transparent analytical expressions may be obtained
from Eqs. (\ref{eq:comp}). In the case $h\neq0$, we obtain
\begin{align}
f_\text{ESE} &= \sum_\sigma [-2\gamma s(\theta)]/\sinh(\lambda_\sigma d),\; f_\text{OSO} = 0,\notag\\
f_\text{ETO} &= 0,\; f_\text{OTE} = \sum_\sigma [-2\sigma\gamma s(\theta)]/\sinh(\lambda_\sigma d)
\end{align}
Note that for $h=0$, $\lambda_\sigma$ becomes independent of
$\sigma$, leading to $f_\text{OTE}=0$. At first glance, this
appears to be in contradiction to Tab. \ref{tab:symmetry} since
the odd-parity components are absent even in the ballistic limit.
However, evaluation of Eqs. (\ref{eq:comp}) for $x\neq d$ reveals
that these components are in general induced, as they should be.
It is remarkable that the odd-parity components vanish exactly
right at the F/I interface. In the presence of a finite exchange
field $h\neq0$, however, the odd-frequency component
$f_\text{OTE}$ survives at $x=d$, and its influence on physical
quantities such as the DOS may be directly probed there. These
results suggest that in order to investigate the influence of the
odd-frequency superconducting correlations $f_\text{ETO}$ and
$f_\text{OSO}$, one would have to measure the DOS at several
positions in the ferromagnetic region and not only at the F/I
interface. In Fig. \ref{fig:hneq0}, we plot the different symmetry
components of the anomalous Green's function in the ferromagnet
and their dependence on the impurity level.

\subsection{Density of states}\label{sec:dos}

\begin{figure}[htb]
\begin{center}
\scalebox{0.4}{
\includegraphics[width=19.0cm,clip]{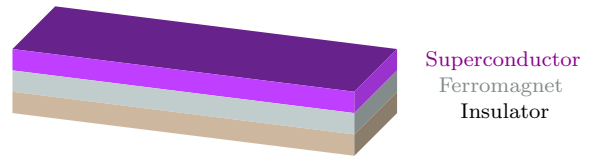}}
\end{center}
\caption{(color online) Setup for our study of the density of states.} \label{fig:model1}
\end{figure}

To demonstrate the applicability of Eq. (\ref{eq:DOS}), we study
in particular how the DOS depends on the crossover from the
ballistic $(g_\text{imp}=0)$ to the diffusive limit ($\gi \gg
\{h,\varepsilon,\Delta_0. \gs\}$). We will fix $\gamma=0.05$ and
$h/\Delta_0=15$ to model a realistic experiment, corresponding
to a weak ferromagnetic alloy like Cu$_{1-x}$Ni$_x$ or
Pd$_{1-x}$Ni$_x$. The setup is shown in Fig. \ref{fig:model1}. It
is well-known that the DOS oscillates in space upon penetration
deeper into the ferromagnetic region \cite{baladie} due to the
presence of an exchange field, a feature which is robust both in
the clean and dirty limit. However, the energy-dependence of the
DOS in the presence of an arbitrary impurity concentration has not
received much attention so far. This is because most works
concerned themselves with the simplified Usadel equation
(diffusive limit) or the Eilenberger equation in the absence of
impurities (clean limit).
\par
In Ref.~\onlinecite{bergeret02}, corrections to the normal-state
DOS as induced by the proximity effect were calculated under the
assumption that $h\tau_\text{imp} \gg 1$. This case corresponds to
a ferromagnet where the exchange field is considerably larger than
the self-energy associated with the impurity scattering. This may
describe either a strong ferromagnet (one must still demand $h\ll
\varepsilon_F$) or a weak ferromagnet with weak impurity
scattering. Neither of these cases are possible to treat with the
Usadel equation. In the present work, however, we do not impose
any restrictions on the parameter $h\tau_\text{imp}$, which allows
us to study the full crossover regime. This may be important in
order to obtain a larger degree of consistency between theory and
experimental data in the case when the diffusive limit is not
fully reached.

\par
In Fig. \ref{fig:DOS_ferromagnet}a), we study the energy-resolved
DOS for an intermediate range of impurity scattering. As a measure
of the junction width, we use the superconducting coherence length
in the clean limit $\xi_S = v_F/\Delta_0$. To isolate the role of
the impurity scattering, we fix the junction width at $d/\xi_S =
0.5$. For a superconductor with $v_F = 10^5$ m/s and $\Delta_0 =
1$ meV, this corresponds to $d \simeq 30$ nm, which is
experimentally relevant. As seen, the DOS exhibits a slightly
oscillating behaviour as a function of energy when the impurity
scattering rate $g_\text{imp}$ is comparable in magnitude to the
superconducting gap. This effect becomes more obvious for wider
junctions $d/\xi_S\gg1$, and is attributed to bound states
appearing in the ferromagnetic film. We discuss this in more
detail below. As $g_\text{imp}$ increases, however, the DOS
becomes featureless for subgap energies although one may still
observe an alternating positive and negative correction to the
zero-energy DOS upon increasing $g_\text{imp}$. In Fig.
\ref{fig:DOS_ferromagnet}b), we plot the spatially-resolved DOS at
$\varepsilon=0$ for various rates of the impurity scattering,
including the case when $h\tau_\text{imp}\sim 1$. As seen, the
oscillations of the zero-energy DOS are reduced with increasing
impurity scattering. We have also investigated the effect of
spin-flip scattering for an intermediate value of the impurity
concentration. The spin-flip scattering, here taken to be
uniaxial, is pair-breaking and thus suppresses the
proximity-effect induced by the superconductor. This aspect agrees
with Ref.~\onlinecite{linder08}, which found that both the triplet
and singlet components are suppressed with uniaxial and/or
isotropic spin-flip scattering. For other types of magnetic
scattering, such as planar spin-flip or spin-orbit scattering, the
singlet and triplet components are affected very differently
\cite{linder08}.

\par
The oscillations of the DOS in S/F junctions are usually
attributed to the oscillating decay of the Cooper pair
wavefunction in the ferromagnetic region. In an S/N junction, this
decay is monotonous, and hence one would not expect to see any
oscillations in the DOS. However, we underline that the impurity
scattering plays an important role in this respect. In the
ballistic case $g\to0$, the proximity of the superconductor
induces Andreev-bound states with well-defined trajectories which
propagate in the normal part of the system. The statistical
distribution of all possible trajectories is peaked at given
lengths, typically at trajectories corresponding to the first and
second reflection processes at the interface. As a result, the DOS
in a \textit{clean} S/N junction acquires oscillations both as a
function of energy and coordinate inside the normal region as seen
in Fig. \ref{fig:DOS_normal} upon averaging over all possible
trajectories. This effect is known as Tomasch-oscillations
\cite{tomasch}.
\par

\begin{widetext}
\text{ }\\
\begin{figure}[htb]
\begin{center}
\scalebox{0.95}{
\includegraphics[width=19.0cm,clip]{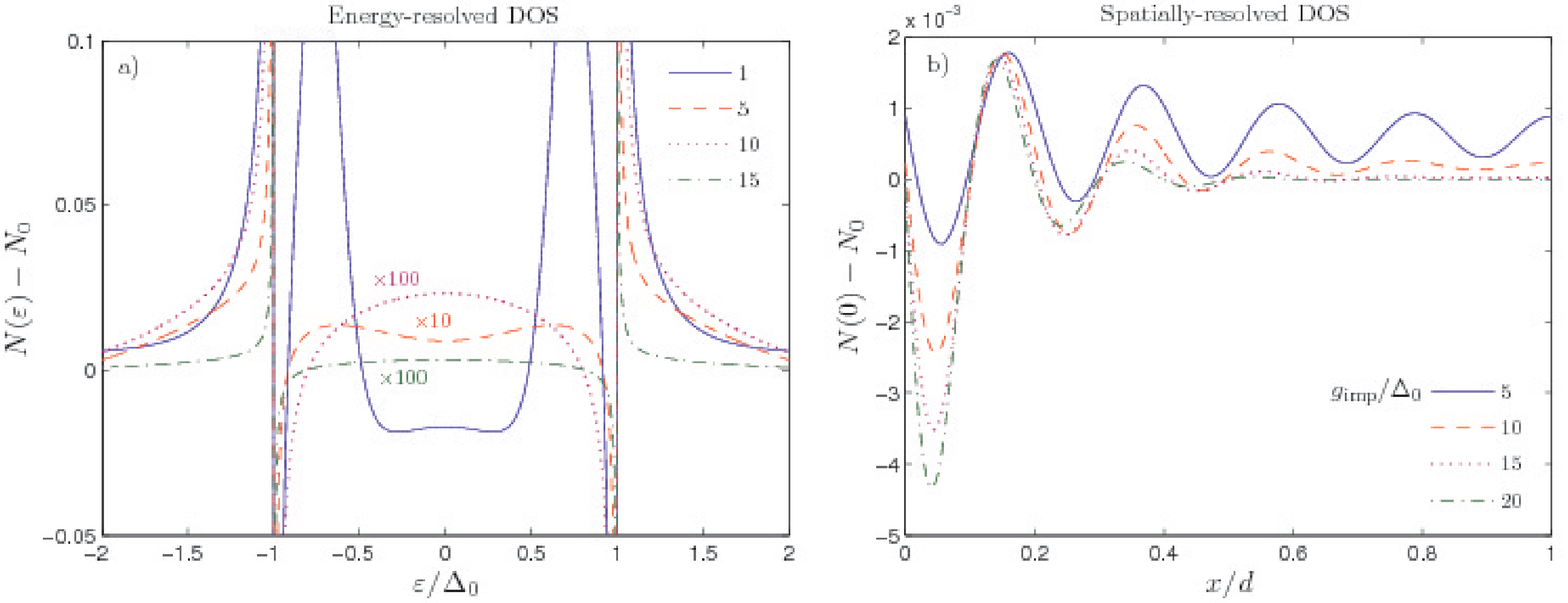}}
\end{center}
\caption{(color online) Plot of the a) energy-resolved DOS at
$x=d$ and b) spatially-resolved DOS at $\varepsilon=0$ for several
values of the impurity scattering rate. Here, the exchange field
is set to $h/\Delta_0=15$ and $d/\xi=0.5$. }
\label{fig:DOS_ferromagnet}
\end{figure}
\text{ }\\
\begin{figure}[htb]
\begin{center}
\scalebox{0.95}{
\includegraphics[width=19.0cm,clip]{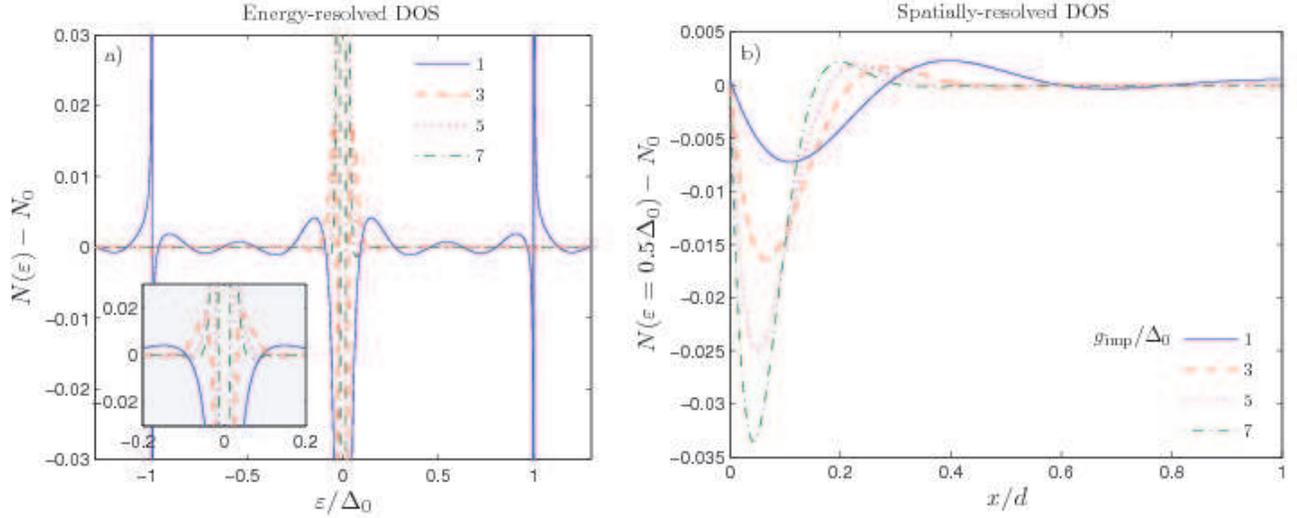}}
\end{center}
\caption{(color online) Plot of the a) energy-resolved DOS at
$x=d$ and b) spatially-resolved DOS at $\varepsilon/\Delta_0=0.5$
for several values of the impurity scattering rate. Here, the
exchange field is set to zero, corresponding to a normal metal,
and $d/\xi=5.0$.} \label{fig:DOS_normal}
\end{figure}
\end{widetext}

\par
However, there is another point which has appears to have been
overlooked in the literature: namely that the spatial oscillations
of the DOS in a S/N junction at finite energies \textit{do not
vanish} in the diffusive limit. Hence, the oscillating DOS as a
function of distance penetrated into the non-superconducting
region is not a feature pertaining uniquely to F/S junctions, as
have been implied in some works \cite{buzdinbrief}. To see this,
we plot the spatially-resolved DOS both for a F/S and N/S junction
in Fig. \ref{fig:snoscillations} in the diffusive regime. The
curves are obtained by using the framework of Ref.
\cite{linder08}, and thus correspond to a full numerical solution
of the Usadel equation without restricting ourselves to the weak
proximity effect regime. The oscillations of the DOS in the N/S
case may be understood by noting that the induced superconducting
Green's function in the normal region has a finite center-of-mass
momentum $q=2\varepsilon/v_F$. This is typically much smaller than
the center-of-mass momentum acquired in a ferromagnet, $q =
2h/v_F$, which means that the corresponding oscillation length is
much larger, but still present. \par Having stated this, it should
be noted that the oscillating nature of the anomalous Green's
function does not necessarily imply that the critical temperature dependence
or the Josephson current in N/S multilayers is non-monotonuous,
\eg displaying $0$-$\pi$ oscillations, since the energy-dependence of the Green's functions is integrated out when obtaining the critical temperature or critical current. For an F/S
junction, on the other hand, the Cooper pair wave-function may
retain its oscillating character even after the energy-integration since the center-of-mass momentum
depends on the exchange field $h$.
\begin{widetext}
\text{ }\\
\begin{figure}[htb]
\begin{center}
\scalebox{0.9}{
\includegraphics[width=19.0cm,clip]{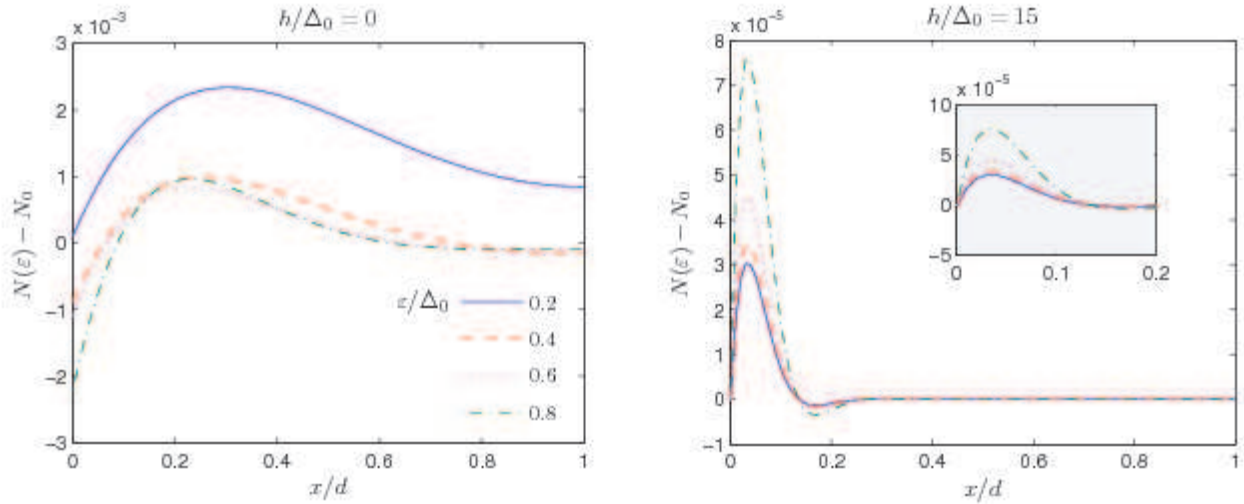}}
\end{center}
\caption{(color online) Plot of the spatially resolved DOS for a
diffusive N/S and F/S junction, respectively. In both cases,
oscillations of the DOS are seen at finite energies. We have here
fixed $d/\xi=3.0$ and $\tau=0.2$, using the notation of Ref.
\cite{linder08} (here, $\xi=\sqrt{D/\Delta_0}$ while $\tau$
denotes the barrier transparency).} \label{fig:snoscillations}
\end{figure}
\end{widetext}

\subsection{Josephson current}\label{sec:jos}
\begin{figure}[htb]
\begin{center}
\scalebox{0.4}{
\includegraphics[width=19.0cm,clip]{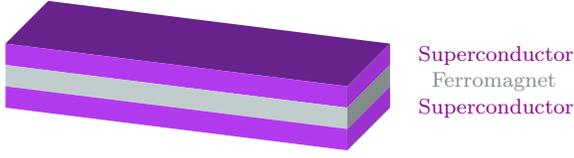}}
\end{center}
\caption{(color online) Setup for our study of the Josephson current.} \label{fig:model2}
\end{figure}
\par
We now evaluate the Josephson current in an SFS junction for an
arbitrary impurity concentration, with a setup as shown in Fig.
\ref{fig:model2}. Denoting the phase at the left (right)
superconductor as $+\chi$ $(-\chi)$, the total phase difference is
given by $\varphi=2\chi$. The current through the junction is
evaluated by
\begin{align}
\mathbf{I}_J = \frac{N_FS_0ev_F}{4} \int \text{d}\varepsilon \tanh(\beta\varepsilon/2)\text{Tr}\{ \langle \hat{\rho}_3 \mathbf{e}_F(\hat{g}^R - \hat{g}^A)\rangle \}
\end{align}
under the assumption of equilibrium distribution functions. Here,
$S_0$ is the effective area of the contact through which the
current flows, while $\beta=1/T$ is inverse temperature.
Experimentally, one measures the current that flows through the
junction, corresponding to the $x$-direction here. We employ the
following boundary conditions:
\begin{align}
\underline{\mathcal{A}_{\sigma,F}}|_{x=0} = \gamma [\mathcal{S}_{\sigma,F}, \mathcal{S}^\text{Left}_{\sigma,S}]_-|_{x=0},\notag\\
\underline{\mathcal{A}_{\sigma,F}}|_{x=d} = -\gamma [\mathcal{S}_{\sigma,F}, \mathcal{S}^\text{Right}_{\sigma,S}]_-|_{x=d},
\end{align}
and approximate $\mathcal{S}_{\sigma,F}=\underline{\tau_3}$ as in
the previous section, in accordance with our assumption of a weak
proximity effect. After some calculations, we arrive at the
following expression for the Josephson current:
\begin{align}\label{eq:josmain}
I_J &= 4\gamma^2N_FS_0ev_F I_c\sin\varphi, \text{ with the definition}\notag\\
I_c &= \int^\infty_{-\infty}\text{d}\varepsilon \sum_\sigma \text{Re}\Bigg\{\frac{s^2(\theta)(1-G_\sigma)\tanh(\beta\varepsilon/2)}{\i(1+G_\sigma)\sinh(\lambda_\sigma d)}\Bigg\}.
\end{align}
The reader is reminded of the definitions
\begin{align}
G_\sigma &= g/(\sqrt{r_\sigma^2 - g^2} + r_\sigma),\notag\\
r_\sigma &= 2\i(\varepsilon+\sigma h) - (\gi + 3\gs)/2,\notag\\
g &= (\gi - \gs)/2,\; g_\text{imp(sf)} \equiv \tau_\text{imp(sf)}^{-1}.
\end{align}
Eq. (\ref{eq:josmain}) is the second of our three main analytical
results in this work. It is probably the most compact way of
expressing the Josephson current for arbitrary exchange fields and
impurity scattering rates within the quasiclassical framework. It
is thus suitable both for the case of a weak ferromagnet (such as
the alloy Cu$_{1-x}$Ni$_x$), and for strong ferromagnets (like Co
or Fe) regardless of whether they are clean or dirty. In experiments performed with
such strong ferromagnets, where the exchange field may be of order
100 meV $(\gg T_c)$, the Usadel equation is not valid at the same
time as the clean limit may not be fully reached. In this case,
one has to use an expression valid for the crossover regime, which
emphasizes the importance of Eq. (\ref{eq:josmain}).

\par
Below, we will study how impurity scattering affects both the
width- and temperature-dependence of the critical current, as well
as its belonging 0-$\pi$ phase diagram. Bergeret \etal
\cite{bergeretstrong} investigated this in the limiting cases of
$h\tau_\text{imp} \ll1$ and $h\tau_\text{imp} \gg1$, while the
majority of studies so far considered exclusively the limiting
case of diffusive motion. We here pay particular attention to the
crossover between the ballistic and diffusive sector, which has
not been investigated previously. To model inelastic scattering,
we add a small imaginary number to the quasiparticle energy,
$\varepsilon\to\varepsilon+\i\delta$ where $\delta=10^{-3}$.
\par
\begin{figure}[htb]
\begin{center}
\scalebox{0.44}{
\includegraphics[width=19.0cm,clip]{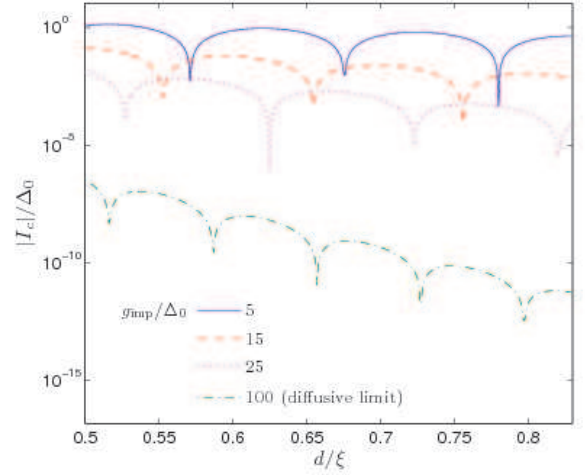}}
\end{center}
\caption{(color online) Plot of the critical current as a function
of junction width $d$. We have used $T/T_c=0.2$.}
\label{fig:josephson_d}
\end{figure}

\begin{figure}[htb]
\begin{center}
\scalebox{0.44}{
\includegraphics[width=19.0cm,clip]{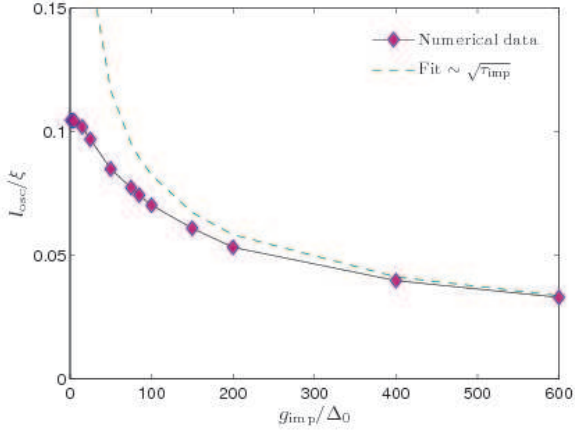}}
\end{center}
\caption{(color online) Plot of the oscillation length of the
critical current as a function of the impurity scattering strength
$g_\text{imp}$. We have used $T/T_c=0.2$.} \label{fig:osclength}
\end{figure}

In Fig. \ref{fig:josephson_d}, we plot the width-dependence of the
critical current for a temperature $T/T_c=0.2$. As seen,
increasing impurity scattering suppresses the magnitude of the
current and also reduces the oscillation length $l_\text{osc}$.
The dependence of the latter on impurity scattering is shown
explicitly in Fig. \ref{fig:osclength}. Using the Usadel equation,
it is predicted that the oscillation length of the critical
current in the dirty limit should depend on the impurity
scattering rate like $\sqrt{h l_\text{imp}} \sim \sqrt{\tau_\text{imp}}$ (for a discussion of the characteristic decay and oscillation lengths in the clean and dirty limit, see Tab. I in Ref. \cite{buzdinrmp}). We obtain a good fit
with this in Fig. \ref{fig:osclength} when $g_\text{imp} \gg
\Delta$. For values of $g_\text{imp}$ comparable to $\Delta$,
however, the oscillation length saturates at a finite value. In
the ballistic limit, the oscillation length is known to depend on
the exchange field like $1/h$. We have also confirmed this for
several values of $h$ when $g_\text{imp} \sim \Delta$.

\par
Also, one notes from Fig. \ref{fig:josephson_d} that the decay
length of the current increases with the concentration of
impurities. It should be noted that the measure $\xi$ used as a
length unit in this context is \textit{independent} of the
impurity scattering rate, since we are using $\xi=v_F/\Delta_0$.
This way, we ensure that the effects observed are really due to
the increased impurity scattering. If we for instance had used the
mean free path $l_\text{mfp} = v_F\tau_\text{imp}$ as a measure
for the junction width, the scale would have been different for
each value of $g_\text{imp}$ in Fig. \ref{fig:josephson_d}. We
also underline that the dirty limit condition is that
$\xi/l_\text{mfp}\gg1$, while the size $d$ of the sample may be
either smaller or larger than $\xi$ as long as that condition is
fulfilled.
\par

\begin{figure}[htb]
\begin{center}
\scalebox{0.44}{
\includegraphics[width=19.0cm,clip]{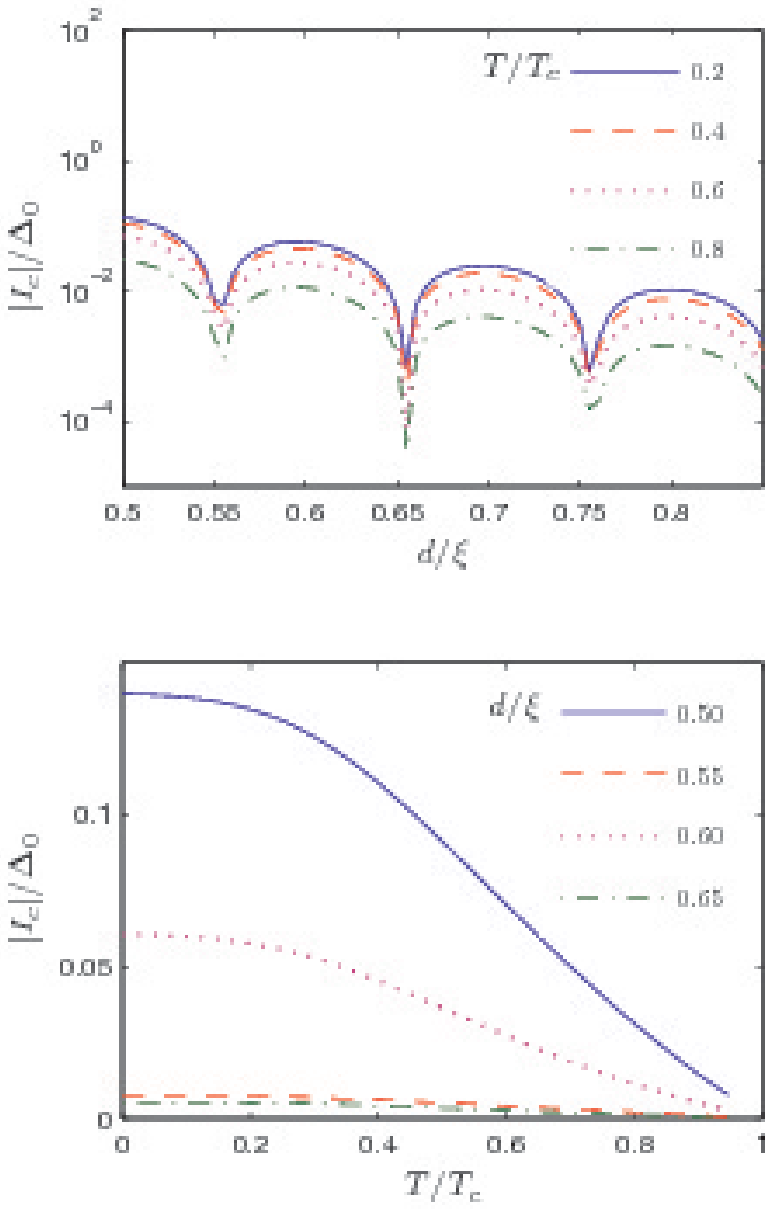}}
\end{center}
\caption{(color online) Plot of the $d$-dependence and the
$T$-dependence of the critical current for
$g_\text{imp}/\Delta_0=15$, corresponding to
$h\tau_\text{imp}=1$.} \label{fig:htau1}
\end{figure}

\begin{figure}[htb]
\begin{center}
\scalebox{0.34}{
\includegraphics[width=19.0cm,clip]{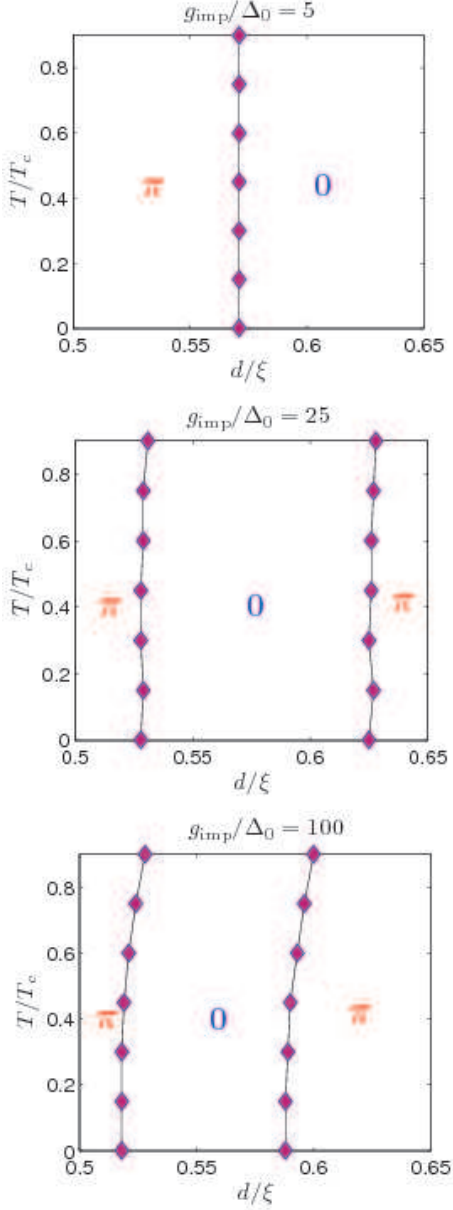}}
\end{center}
\caption{(color online) Phase diagram in the $d$-$T$ plane for the
$0$-$\pi$ transitions of the critical current for several values
of the impurity concentration.} \label{fig:Phasediagram}
\end{figure}

In Fig. \ref{fig:htau1}, we pay particular attention to the case
$h\tau_\text{imp} = 1$ which is inaccessible in the Usadel
framework. As seen, nothing qualitatively new shows up in the
$d$-dependence or the $T$-dependence of the critical current as
compared to the diffusive limit, although the decay rate is
considerably lower. We also investigate how the $0$-$\pi$ phase
diagram of the Josephson junction is affected by impurity
scattering. This is most conveniently plotted in the $d$-$T$
plane. In Fig. \ref{fig:Phasediagram}, one observes several
features. First, it is clear that the area occupied by the $0$ and
$\pi$ phases, respectively, diminishes with increasing
$g_\text{imp}$, in agreement with the shortened oscillation length
of Fig. \ref{fig:osclength}. Secondly, it is seen that thermal
$0$-$\pi$ transitions are practically speaking impossible to
observe for scattering rates satisfying $g_\text{imp} \leq h$. As
the scattering rate is increased, however, the thermal transitions
become possible when $g_\text{imp} \gg h$, or equivalently
$h\tau_\text{imp} \ll 1$. In this regime, the Usadel equation is
valid and we obtain consistency with previous results. At all
scattering rates, the width-induced transitions are possible.

\subsection{Critical temperature}\label{sec:tc}
\begin{figure}[htb]
\begin{center}
\scalebox{0.4}{
\includegraphics[width=19.0cm,clip]{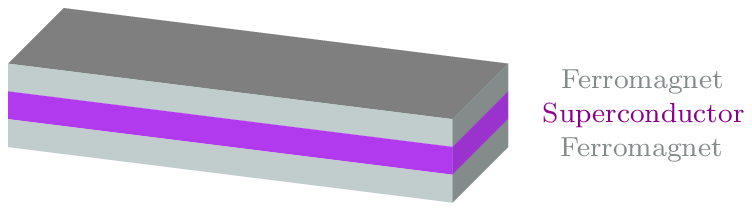}}
\end{center}
\caption{(color online) Setup for our study of the critical temperature.} \label{fig:model3}
\end{figure}

Finally, we investigate an F/S/F layers where the critical
temperature of the superconductor is sensitive to the relative
orientation of magnetization of the two F layers. This effect is
usually dubbed to a \textit{spin-switch} effect in the literature.
Our setup is shown in Fig. \ref{fig:model3}. Tagirov
\cite{tagirov} was the first to point out the interesting
opportunity to "activate" superconductivity simply by means of
switching the direction of the magnetization in one of the
ferromagnetic layers. Since then, a number of works have
elaborated on the spin-switch effect both experimentally
\cite{tc1,tc2,tc3} and theoretically
\cite{buzdin_epl_99, baladie,fominov,eschrigTC}. In particular, a convincing
numerical approach was developed in Ref. \cite{fominov}. So far,
however, almost all theoretical works focused on the dirty limit,
in which the critical temperature may be conveniently calculated
by using the Usadel equation in the Matsubara frequency
representation. Although the obtained results compare well
qualitatively with experimental data, an unsolved factor so far is
the discrepancy of two orders in magnitude of the predicted
effect. Recently, it was proposed and investigated \cite{zamensky}
if an asymmetry in the interface transparencies of the F/S/F
junctions could be responsible for this, in effect one of the
interfaces was much less transparent than the other. The authors
of Ref. \cite{zamensky} concluded that this was not the case. At present, the single
ferromagnet F/S/F devices to have been examined so far have used
strong ferromagnets, which falls outside the range of
applicability of the Usadel equation \cite{tc1,tc2,tc3}. In light
of this, it would be interesting to go beyond the usual treatment
with the Usadel equation and solve the more general Eilenberger
equation to investigate the role of the impurity scattering.
\par
A general analytical solution for arbitrary proximity effect and
barrier transparency is hardly achievable, as pointed out
previously. Nevertheless, it is reasonable to expect that one may
capture the essential physics in the weak proximity effect regime.
In order to calculate the critical temperature for the P and AP
alignment, we assume that the temperature is close to $T_c$, which
allows us to write the Green's function in the superconductor as
follows:
\begin{align}
\underline{g_\sigma} = \begin{pmatrix}
1 & 2a_\sigma \\
2b_\sigma & - 1\\
\end{pmatrix},
\end{align}
since $\lim_{\Delta\to 0} c(\theta) = 1$. For the normal part of
the Green's function matrix, this means that $(1-a_\sigma
b_\sigma)/(1+a_\sigma b_\sigma) \simeq 1$, while for the anomalous
Green's function one thus has $2a_\sigma/(1+a_\sigma b_\sigma)
\simeq 2a_\sigma$. The self-consistency equation for the
superconducting gap reads in general \cite{jpdiplom}
\begin{align}
\Delta = \frac{N_F\lambda}{8} \text{Tr}\Big\{ \Big( \frac{\hat{\rho}_1 - \i\hat{\rho}_2}{2} \Big) \hat{\tau}_3 \int \text{d}\varepsilon \langle \hat{g}^K \rangle\Big\},\; \lambda>0,
\end{align}
where $\lambda$ is the attractive interaction and $\hat{g}^K$ is
the Keldysh part of the Green's function. For an equilibrium
situation $[\hat{g}^\text{K} = (\hat{g}^\text{R} -
\hat{g}^\text{A})\tanh(\beta\varepsilon/2)]$ in the weak-proximity
effect regime with a temperature very close to $T_c$, this reduces
to
\begin{align}
\Delta = \frac{N_F\lambda}{8} \int \text{d}\varepsilon \tanh\Big(\frac{\varepsilon}{2T_c}\Big) \sum_\pm \sum_\sigma \sigma[a_\sigma^\pm - (b_\sigma^\pm)^*].
\end{align}
Once the anomalous Green's functions
$\{a_\sigma^\pm,b_\sigma^\pm\}$ have been obtained, one may solve
Eq. (\ref{eq:Tc}) numerically to obtain $T_c$ in the P and AP
configurations. Using boundary conditions explained below, we
solve for the anomalous Green's functions in both the
ferromagnetic and superconducting regions and obtain the following
equation determining the critical temperature:
\begin{widetext}
\begin{align}\label{eq:Tc}
1 &- N_F\lambda \int^\omega_0 \text{d}\varepsilon \tanh\Big(\frac{\varepsilon}{2T_c}\Big)\varepsilon^{-1}\Big[1-\cos(2\varepsilon x/v_F) - \sum_{\sigma,\pm} \text{Re}\Big\{
\frac{L_\sigma^\pm \e{\pm2\i\varepsilon x/v_F} \sum_\alpha  \alpha R_\sigma^\alpha(1-\e{2\alpha\i\varepsilon d_S/v_F})}{4\sum_\alpha \alpha\e{2\alpha\i\varepsilon d_S/v_F}L_\sigma^\alpha R_\sigma^\alpha } \Big\}\Big] = 0,
\end{align}
\end{widetext}
with the cut-off energy $\omega$, $\alpha=\pm$, and finally
\begin{align}
L_\sigma^\pm &= \e{\pm \lambda_\sigma^\text{Left} d_F} - G_\sigma^\text{Left}\e{\mp \lambda_\sigma^\text{Left} d_F},\notag\\
R_\sigma^\pm &= \e{\pm \lambda_\sigma^\text{Right} d_F} - G_\sigma^\text{Right}\e{\mp \lambda_\sigma^\text{Right} d_F}.
\end{align}
Eq. (\ref{eq:Tc}) is the third of our three main analytical
results in this work. It gives an expression for the critical
temperature in an F/S/F junction for arbitrary exchange fields and
impurity scattering rates within the framework of quasiclassical
theory in the weak-proximity effect regime.
\par
In order to find $\{a_\sigma^\pm,b_\sigma^\pm\}$, we must
introduce proper boundary conditions at each of the interfaces in
the setup (Fig. \ref{fig:model3}). The left ferromagnet is assumed
to occupy the region $-d_F<x<0$, the superconductor is located at
$0<x<d_S$, while the right ferromagnet occupies the space
$d_S<x<d_S+d_F$. Thus, the ferromagnetic layers are assumed to
have the same thickness $d_F$ while the superconductor has
thickness $d_S$. Due to the complexity of the problem, we will
assume rigid boundary conditions at the superconductor/ferromagnet
interfaces, which amounts to continuity of the Green's function.
Although the low transparency limit is probably more realistic, it
is reasonable to expect qualitatively correct results in this
approach. Moreover, since we already assume a temperature close to
$T_c$, the proximity effect would be almost completely absent if
we in addition incorporated tunneling interfaces. In general, high 
transparency interfaces cause a depletion of the superconducting order
parameter near the interface, which means that one should strictly speaking
solve for the spatial depletion of the gap self-consistently. In our approach,
we do not incorporate this depletion since we are aiming for analytical results. 
A full numerical approach would, however, doubtlessly improve the accuracy of the results presented
below, but at the prize of losing the analytical information.
\par
At the ends of
the ferromagnetic layers, we impose vacuum boundary conditions. In
total, the boundary conditions then read:
\begin{align}
x=-d_F:&\hspace{0.2in} \underline{\mathcal{A}_{\sigma,F}^\text{Left}} = \underline{0},\notag\\
x=0:&\hspace{0.2in} \underline{g_{\sigma,F}^\text{Left}} = \underline{g_\sigma},\notag\\
x=d_S:&\hspace{0.2in} \underline{g_{\sigma}} = \underline{g_{\sigma,F}^\text{Right}},\notag\\
x=d_S+d_F:&\hspace{0.2in} \underline{\mathcal{A}_{\sigma,F}^\text{Right}} = \underline{0},\notag\\
\end{align}
After straight-forward calculations, we obtain an expression for
$\{a_\sigma^\pm,b_\sigma^\pm\}$ in the superconductor. A few
comments with regard to the expression Eq. (\ref{eq:Tc}) are in
order. Firstly, it should be noted that the expression for the
critical temperature in Eq. (\ref{eq:Tc}) depends on the position
$x$ in the superconductor through the spatial dependence of the
anomalous Green's function. This dependence is of course
artificial and a result of the approximations we have made in the
calculations; in a real experimental sample, $T_c$ is a property
for the entire layer and does not depend on the position in the
superconductor. The reason for why we obtain an artificial
$x$-dependence in the expression for the critical temperature is
because we have neglected the spatial modification of the order
parameter $\Delta$ in the layer. Employing a fully self-consistent
calculation would remove the spatial dependence of $T_c$ in the
gap equation. However, for thin superconducting layers $d_S/\xi
\ll 1$, our approximation is expected to be good. A similar
procedure has been used in several other works which calculated
$T_c$ by means of the Usadel equation. In those works, it was
assumed that the anomalous Green's function in the superconductor
varied very little as long as $d_S/\xi \ll 1$ was satisfied, and
hence one could ignore the spatial dependence of the Green's
function once it had been found. More precisely, $T_c$ was
evaluated in the middle of the superconducting region. In our
case, we will use the same approximation since our approach is
analytical in nature. The main contribution to the integral in Eq.
(\ref{eq:Tc}) comes from energies $\varepsilon \leq \Delta$, for
which the terms including the coordinate $x$ on the right hand
side of the equation change very little as long as $d_S/\xi \ll
1$. We will focus on the
difference between the critical temperature in the P and AP
alignments, defined as
\begin{equation}
\Delta T_c \equiv T_c^\text{AP} - T_c^\text{P}.
\end{equation}
We will normalize all temperatures on $T_c^0$, which is the bulk
critical temperature of the superconductor in the absence of a
proximity effect. As demanded by consistency, the critical
temperature approaches $T_c^0$ when $d_F\to0$. We choose the cut-off frequency as $\omega/\Delta_0=30$.
\par
\begin{widetext}
\text{ }
\begin{figure}[htb]
\begin{center}
\scalebox{0.9}{
\includegraphics[width=19.0cm,clip]{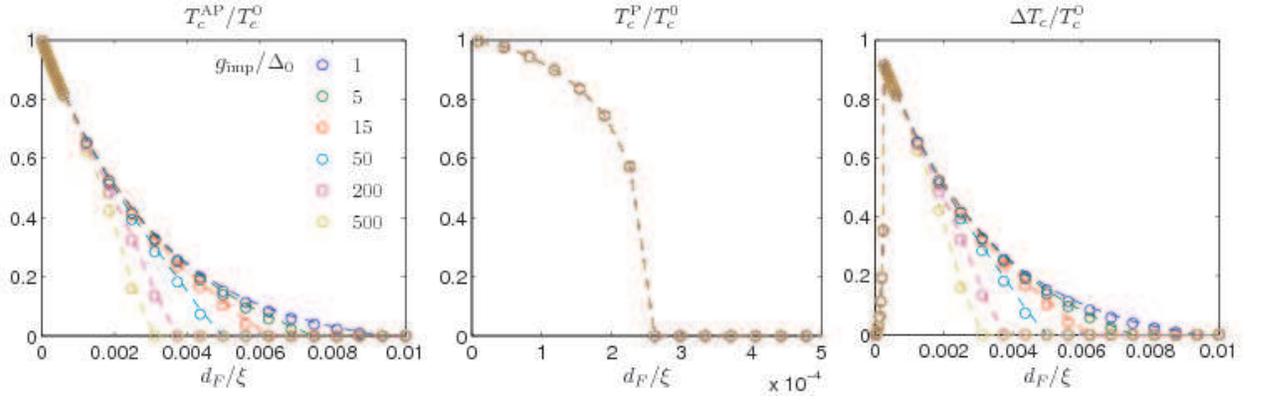}}
\end{center}
\caption{(color online) Plot of the critical temperature in an
F/S/F structure as a function of the ferromagnetic layer thickness
$d_F$ for fixed $d_S/\xi=0.03$. Note the different scale for $d_F$ in the middle panel. The symbols denote the result for $x/d_S=0.50$ while the dashed lines correspond to $x/d_S=0.01$.}
\label{fig:d003}
\end{figure}
\end{widetext}
With the analytical solution in hand, we now present a study of
the critical temperature in the P and AP configuration,
investigating in particular the role of impurity scattering.
First, we plot the critical temperature as a function of
ferromagnetic layer thickness with a fixed superconducting layer
thickness of $d_S/\xi=0.03$ in Fig. \ref{fig:d003}. Using a superconductor with $\xi=200$ nm, this would correspond to a thickness $d_S=6$ nm. To ensure the validity of our assumption that the anomalous Green's functions vary little with $x$ throughout the superconducting layer, we plot the critical temperature both at $x/d_S=0.50$ (symbols) and $x/d_S=0.01$ (dashed lines). As seen, the difference is neglible. From Fig. \ref{fig:d003}, one may infer that the critical temperature in the P configuration goes to zero much faster than in the AP configuration as a function of the ferromagnetic layer thickness $d_F$. This supports the notion that the antiparallell configuration favors superconductivity in the middle layer. The effect of impurity scattering is seen to suppress the critical temperature, in general. 
\par
One may understand intuitively why the antiparallell alignment is favorable compared to the parallell alignment, since the average exchange field cancels in the former case. Qualitatively, our results are consistent with the monotonic decay found for a high barrier transparency when using the Usadel equation \cite{buzdinrmp}. However, a more realistic scenario would clearly be to invoke low barrier transparency boundary conditions at the S/F interfaces. Due to the complexity of the problem upon including an arbitrary amount of impurities, we have used perfectly transparent interfaces here as a first approximation. It would nevertheless be quite interesting to extend this formalism to low transparency interfaces to investigate the role of impurity scattering under those circumstances. Especially, the role of $g_\text{imp}$ with regard to the re-entrant behavior of $T_c$ would be worth investigating. Our analytical results may serve as a basis for extending this formalism to low transparency interfaces in the case of an arbitrary value for $h\tau_\text{imp}$, as opposed to $h\tau_\text{imp}$ in the Usadel regime.

\section{Summary}\label{sec:summary}
We have investigated various aspects of the physics resulting from
the proximity effect in ferromagnet/superconductor (F/S) bilayers.
In contrast to previous works, which were  limited to either the
clean or dirty limit, we have taken into account an arbitrary
scattering rate for both non-magnetic and magnetic impurities.
This has allowed us to access the crossover regime from the
ballistic to diffusive regime of the proximity effect. We have
derived analytical formula for \textit{i)} the proximity-induced
DOS of an F/S bilayer, \textit{ii)} the Josephson current in an
S/F/S junction, and \textit{iii)} the critical temperature of an
F/S/F structure. Our results are valid for an arbitrary ratio of
the parameter $h\tau_\text{imp}$, and are thus applicable both to
weak ferromagnetic alloys as well as permalloys in either the
diffusive or clean limit.

\acknowledgments J. L. acknowledges T. Yokoyama, A. Cottet, F. S.
Bergeret, and Ya. Fominov for useful discussions. H. Skadsem and
M. Thaule are also thanked for valuable input. J.L. and A.S. were
supported by the Norwegian Research Council Grant Nos. 158518/431,
158547/431, (NANOMAT), and 167498/V30 (STORFORSK). M. Z. thanks A.
Brataas and A. Sudb{\o} for their hospitality and support during
his visit to the Centre for Advanced Study, Oslo.

\appendix

\section*{Appendix}
\noindent
The Pauli-matrices used in this paper are defined as
\begin{align}\label{eq:matricesapp}
\underline{\tau_1} &= \begin{pmatrix}
0 & 1\\
1 & 0\\
\end{pmatrix},\;
\underline{\tau_2} = \begin{pmatrix}
0 & -\i\\
\i & 0\\
\end{pmatrix},\;
\underline{\tau_3} = \begin{pmatrix}
1& 0\\
0& -1\\
\end{pmatrix},\notag\\
\underline{1} &= \begin{pmatrix}
1 & 0\\
0 & 1\\
\end{pmatrix},\;
\hat{1} = \begin{pmatrix}
\underline{1} & \underline{0} \\
\underline{0} & \underline{1} \\
\end{pmatrix},\;
\hat{\tau}_i = \begin{pmatrix}
\underline{\tau_i} & \underline{0}\\
\underline{0} & \underline{\tau_i} \\
\end{pmatrix},\notag\\
\hat{\rho}_1 &= \begin{pmatrix}
\underline{0} & \underline{\tau_1}\\
\underline{\tau_1} & \underline{0} \\
\end{pmatrix},\;
\hat{\rho}_2 =  \begin{pmatrix}
\underline{0} & -\i\underline{\tau_1}\\
\i\underline{\tau_1} & \underline{0} \\
\end{pmatrix},\;
\hat{\rho}_3 = \begin{pmatrix}
\underline{1} & \underline{0}\\
\underline{0} & -\underline{1}  \\
\end{pmatrix}.
\end{align}

\end{document}